\documentclass{aa}
\usepackage{graphicx}
\usepackage{txfonts}
\usepackage{siunitx}
\usepackage{hyperref}
\usepackage{lipsum}
\usepackage{bm}
\usepackage{xcolor}

\newcommand{\rmn}[1]{\mathrm{#1}}
\newcommand{\bnabla}{\bm{\nabla}}
\newcommand{\bcdot}{\bm{\cdot}}
\defcitealias{Perrone2022}{PL22a}
\defcitealias{Perrone2022a}{PL22b}
\defcitealias{Perrone2023b}{PBP23}

\usepackage{xpatch}
\usepackage{hyperref}
\hypersetup{
        colorlinks=true,
        breaklinks=true,
        citecolor=blue,
        allcolors=blue,
        frenchlinks=true
}

\makeatletter
\xpatchcmd\NAT@citex
{%
        \@citea\NAT@hyper@{%
                \NAT@nmfmt{\NAT@nm}%
                \hyper@natlinkbreak{\NAT@aysep\NAT@spacechar}{\@citeb\@extra@b@citeb}%
                \NAT@date
        }%
}
{%
        \@citea
        \NAT@nmfmt{\NAT@nm}%
        \NAT@aysep\NAT@spacechar
        \NAT@hyper@{\NAT@date}%
}
{}{}
\xpatchcmd\NAT@citex
{%
        \@citea\NAT@hyper@{%
                \NAT@nmfmt{\NAT@nm}%
                \hyper@natlinkbreak{\NAT@spacechar\NAT@@open\if*#1*\else#1\NAT@spacechar\fi}%
                {\@citeb\@extra@b@citeb}%
                \NAT@date
        }%
}
{
        \@citea
        \NAT@nmfmt{\NAT@nm}%
        \NAT@spacechar\NAT@@open\if*#1*\else#1\NAT@spacechar\fi
        \NAT@hyper@{\NAT@date}%
}
{}{}
\makeatother

\makeatletter
\renewcommand*\aa@pageof{, page \thepage{} of \pageref*{LastPage}}
\makeatother

\begin{document}

\title{Does the magnetothermal instability survive whistler suppression of thermal conductivity in galaxy clusters?}
\titlerunning{The impact of whistler waves on the magnetothermal instability}

        \author{Lorenzo Maria Perrone\inst{1}
                \and
                Thomas Berlok\inst{2,1}
                \and
                Christoph Pfrommer\inst{1}
        }
        
        \institute{Leibniz-Institut f\"{u}r Astrophysik Potsdam (AIP), 
                An der Sternwarte 16, D-14482 Potsdam, Germany\\
                \email{lperrone@aip.de}
      \and
           Niels Bohr Institute, University of Copenhagen, Blegdamsvej 17, 2100 Copenhagen, Denmark
        }
        
        \date{\today}
        
        
        \abstract{
    The hot and dilute intracluster medium (ICM) plays a central role in many key processes that shape galaxy clusters.
    Nevertheless, the nature of plasma turbulence and particle transport in the ICM remain poorly understood, and quantifying the effect of kinetic plasma instabilities on the macroscopic dynamics represents an outstanding problem.
    Here we focus on the impact of whistler-wave suppression of the heat flux on the magnetothermal instability (MTI), which is expected to drive significant turbulent motions in the periphery of galaxy clusters.
    We perform small-scale Boussinesq simulations with a sub-grid closure for the thermal diffusivity in the regime of whistler-wave suppression. Our model is characterized by a single parameter that quantifies the collisionality of the ICM on the astrophysical scales of interest that we tune to explore a range appropriate for the periphery of galaxy clusters. We find that the MTI is qualitatively unchanged for weak whistler suppression. Conversely, with strong suppression the magnetic dynamo is interrupted and MTI turbulence dies out. In the astrophysically relevant limit, however, the MTI is likely to be supplemented by additional sources of turbulence. Investigating this scenario, we show that the inclusion of external forcing has a beneficial impact and revives even MTI simulations with strong whistler suppression. As a result, the plasma remains buoyantly unstable, with important consequences for turbulent mixing in the ICM.
}

        
        \maketitle
        %
        
        \section{Introduction}
        
        Understanding the physics of transport processes in the hot dilute intracluster medium (ICM) of galaxy clusters is an ongoing theoretical challenge. 
        This has important implications for the properties of turbulence and the feedback processes \citep[including from active galactic nuclei,][]{Balbus2008a,Sharma2009b,Reynolds2015a,Yang2016a}. One major complication arises from the fact that high-$\beta$, weakly collisional plasmas are susceptible to kinetic microinstabilities that grow on scales observationally unresolved \citep{Schekochihin2006,Kunz2014}. These instabilities can significantly alter particle motions and can  affect thermal transport \citep{Riquelme2016,Komarov2016,Komarov2018}, plasma viscosity \citep{Zhuravleva2019}, particle acceleration \citep{Sironi2015b}, and magnetic field amplification \citep{St-Onge2020}.
        In the presence of a heat flux (generated, e.g., by a temperature gradient), the whistler instability grows and saturates by regulating the scattering of electrons via magnetic fluctuations, reducing their speed along magnetic field lines to values well below the thermal velocity \citep{Levinson1992,Pistinner1998}. As a result the heat flux can be significantly suppressed below the collisional (Spitzer) value, depending on the plasma $\beta$ \citep{Riquelme2016,Roberg-Clark2016,Roberg-Clark2018,Komarov2018,Drake2021}. Possible consequences of this suppression are far-reaching, and include modifications to the Field length \citep{Drake2021} and inhibition of plasma instabilities  that rely on efficient thermal conduction along magnetic fields, such as the magnetothermal instability \citep[MTI,][]{Balbus2000,Kunz2011}. This instability may be active in the periphery of galaxy clusters and may contribute to the observed levels of turbulence \citep{Parrish2007,McCourt2011}.
        
        In previous work it was shown that the MTI produces a state of turbulence sustained by the background temperature gradient and composed of density and velocity fluctuations over a wide range of scales, with rms values at saturation that follow clear power laws with the thermal diffusivity and the gradients of gas entropy and temperature \citep[][hereafter \citetalias{Perrone2022,Perrone2022a}, respectively]{Perrone2022,Perrone2022a}. Through these scaling laws it was estimated that the MTI is capable of driving turbulent motions of hundreds of  kilometers per second, on scales of $100 \, \si{kpc}$, roughly consistent with existing measurements of turbulence (\citealt{HitomiCollaboration2018}; see \citealt{Kempf2023} for a detailed discussion on the observational constraints on MTI turbulence).
        
        One of the main uncertainties of these estimates is our still limited understanding of the impact of microscale instabilities on the macroscopic heat conduction. A self-consistent model of this impact would require plasma kinetic simulations spanning these very disparate scales. Because of the severe computational costs involved,    a popular approach is to instead adopt sub-grid models for the transport coefficient informed by kinetic theory and simulations \citep{Sharma2006,Kunz2012,Foucart2017,Kempski2019a,Beckmann2022,Squire2023}. 

        Regarding the MTI, studies of its nonlinear saturation with suppression of thermal conduction by the mirror instability revealed very little difference compared to the unsuppressed case \citep{Berlok2021}. On the other hand, the impact of electron scattering at whistler waves could potentially be more severe \citep{Drake2021} and remains unexplored. 
        In this work, and in its companion paper \citep[][hereafter \citetalias{Perrone2023b}]{Perrone2023b}, we address this issue by investigating the survival of the MTI when the plasma is subject to the expected  whistler suppression of its heat conductivity. We do this using controlled small-scale simulations. We show that the MTI survives unscathed for the mild levels of suppression expected in the periphery of galaxy clusters. With stronger suppression, the MTI instead enters a ``spiral-of-death'' scenario. Nevertheless, we show that the presence of external forcing (the result of, e.g., mergers or substructure accretion) revives the MTI rather than further obstructing it. Our findings support the idea that the MTI could be active in the periphery of galaxy clusters even with suppression of thermal conduction.

        \begin{figure*}
                \centering
                \includegraphics[width=1.0\linewidth]{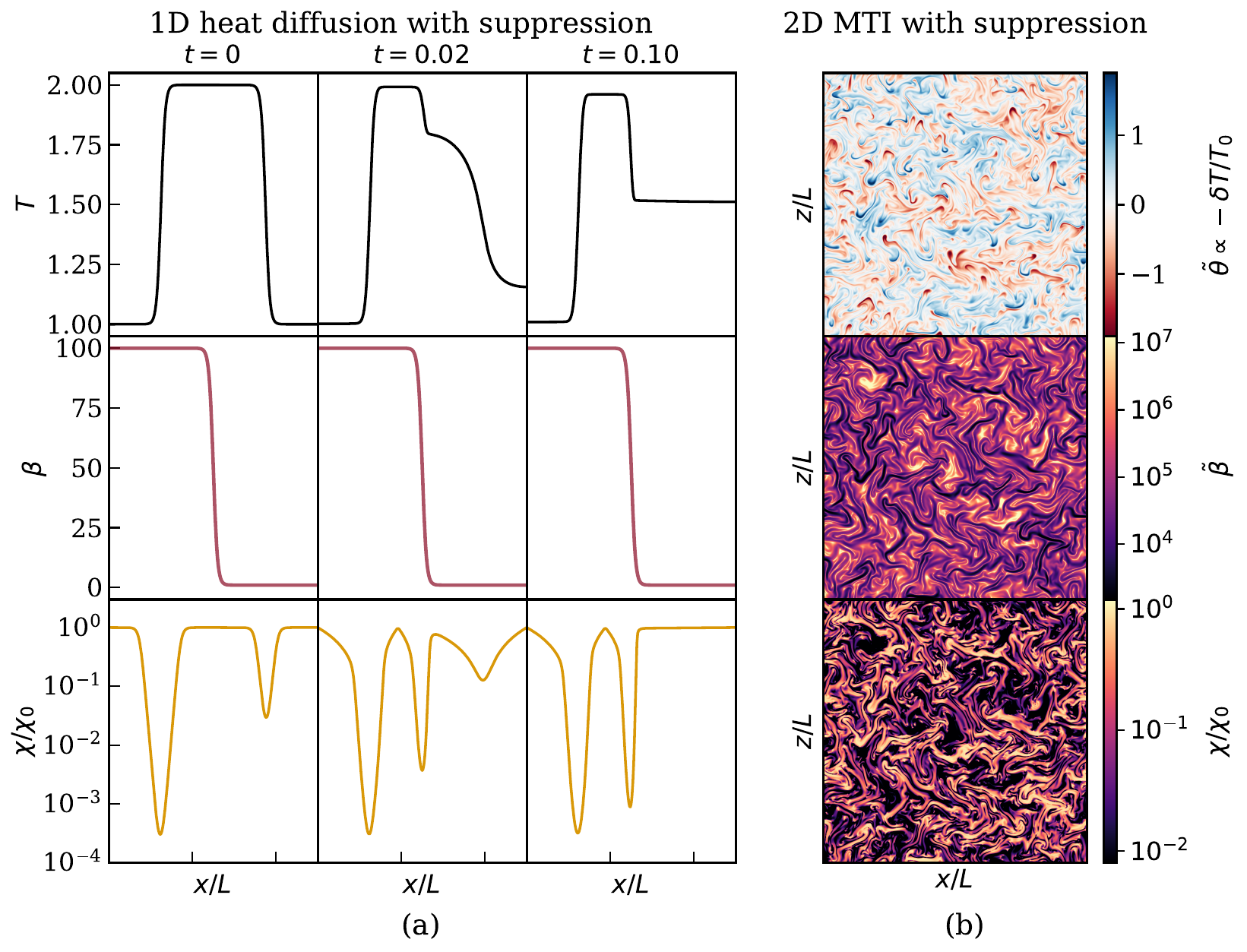}
                \caption{Implementation of whistler suppression in 1D and 2D with the MTI. Part (a): 1D simulation illustrating the behavior of  whistler-suppressed thermal conduction. The temperature (top), plasma $\beta$ (middle, kept fixed), and thermal conductivity (bottom) are shown at different times (in units of $L^2/\chi_0$, with $L$ the box size). For this 1D example the suppression factor has the form $\chi / \chi_0 = (1 + L \beta \mathrm{d} \ln T / \mathrm{d} x )^{-1}$ such that heat conduction is suppressed at strong temperature gradients and high $\beta$. Temperature equilibration thus occurs faster on the right (where $\beta$ is low) than on the left (where $\beta$ is high).  Part (b): Snapshots at saturation of 2D MTI turbulence with a similar form of whistler suppression (Eq.~\ref{eq:thermal_diffusivity}). Shown are  temperature fluctuations (top), the Boussinesq equivalent of plasma beta $\tilde{\beta}$ (middle), and the spatially varying heat conductivity $\chi$ (bottom). The suppression of $\chi$ is seen to spatially correlate with regions of high $\tilde{\beta}$. The MTI-driven temperature fluctuations reveal that the MTI continues to work despite the whistler suppression of the thermal diffusivity, which amounts to $20\%$ of $\chi_0$ (volume average).}
                \label{fig:whistler_suppression_1D_2D}
        \end{figure*}
 
        \section{Model}
        
        We performed local simulations of the MTI with the pseudospectral code \textsc{snoopy} \citep{Lesur2015}. As was done in  \citetalias{Perrone2022}, we adopted the Boussinesq approximation to study small-scale subsonic turbulence, assuming a constant negative background temperature gradient (temperature decreases radially) and a positive entropy stratification that is stable in the absence of anisotropic heat conduction. The two gradients define respectively the MTI ($\omega_{\rmn{T}}$) and Brunt-V\"ais\"al\"a frequency ($N$; see Appendix~\ref{appA}), whose values in the periphery of galaxy clusters are roughly comparable and are approximately $(600 \,\si{Myr})^{-1}$ \citepalias{Perrone2022}. In our local Cartesian setup, the radial direction coincides with $\bm e_z$. The full magnetohydrodynamic (MHD) Boussinesq equations with anisotropic thermal conduction can be found in Appendix~\ref{appA}.

\subsection{Whistler suppression of thermal conduction}
        We employed the following closure for the parallel heat flux in the regime of electron scattering by whistler waves \citep{Komarov2018}:
        \begin{align}
                q_{\parallel} = \frac{q_{\parallel,\rmn{s}}}{1 + \frac{1}{3} \beta_\rmn{e} \lambda_{\text{mfp,e}}/L_{\rmn{T},\parallel}  }. \label{eq:whistler_suppression_approx}
        \end{align}
    Here $\beta_\rmn{e}$ is the electron plasma beta (hereafter $\beta$), $\lambda_{\text{mfp,e}}$ is the electron mean-free path, and $L_{\rmn{T},\parallel}$ is the temperature scale along the magnetic field. In the periphery of galaxy clusters typical values are $\beta \sim 100$, $\lambda_{\text{mfp,e}} \sim 10 \, \si{kpc}$, and $L_{\rmn{T},\parallel} \sim 100 \, \si{kpc}$. Equation~\eqref{eq:whistler_suppression_approx} smoothly interpolates between the Spitzer collisional heat flux \citep[$q_{\parallel,\rmn{s}} \simeq 0.5 n_\rmn{e} m_\rmn{e} \varv_{\text{th,e}}^3 \lambda_{\text{mfp,e}}/L_{\rmn{T},\parallel}$;][where $n_\rmn{e}$ is the electron number density, $m_\rmn{e}$ is the electron mass, and $\varv_{\text{th,e}}$ is the electron thermal speed]{Spitzer1962,Cowie1977} at low $\beta$, and the marginal heat flux controlled by the whistler instability at saturation when $\beta$ is high \citep[$q_{\parallel,\rmn{w}} \simeq 1.5 \beta^{-1} n_\rmn{e} m_\rmn{e} \varv_{\text{th,e}}^3$;][]{Komarov2018}.
        Since in the periphery of galaxy clusters $\lambda_{\text{mfp,e}} / L_{\rmn{T},\parallel} \ll 1$, in our model we do not include  saturation of the heat flux when the electron mean-free path is larger than the temperature scale \citep{Cowie1977}.

        We show the principle of whistler suppression of thermal conductivity through a simplified 1D simulation of thermal diffusion (Fig.~\ref{fig:whistler_suppression_1D_2D}a), where we evolve the temperature equation with anisotropic diffusion in the presence of a fixed background magnetic field that has a spatially varying field strength.  
        As time progresses, sharp temperature gradients survive where $\beta$ is high, but the gradients are gradually erased where the magnetic field is stronger. On the other hand, steep temperature gradients can be generated at the interface between regions of high and low $\beta$ where none was present before (see Fig.~\ref{fig:whistler_suppression_1D_2D}a). 

\subsection{Whistler suppression with Boussinesq}
        To model whistler suppression of the heat flux in the Boussinesq approximation, we introduce the analogue of the usual plasma beta $\tilde{\beta}$ as 
    \begin{align} 
       \tilde{\beta}= \frac{\varv_{\mathrm{cond}}^2}{B^2 / (8 \pi \rho)}, \label{eq:beta_tilde_def}
    \end{align}
    (where $\rho$ is the mass density of the plasma), which we define through the  conduction speed $\varv_{\mathrm{cond}}$ rather than the thermal speed since the latter is ordered out of the Boussinesq equations. The conduction speed is given by $\varv_{\mathrm{cond}} = (\chi_0 \omega_{\rmn{T}})^{1/2}$, where $\chi_0$ is the Spitzer diffusivity. In typical galaxy clusters, $\varv_{\mathrm{cond}} \simeq 500 \, \si{km.s^{-1}}$ and $\tilde{\beta} \simeq 0.5 \beta$. With this substitution, the whistler-suppressed thermal diffusivity associated with Eq.~\eqref{eq:whistler_suppression_approx} can be written as (see Appendix~\ref{appB})
    \begin{align}
                \chi = \frac{\chi_0}{1 + \frac{1}{3} \tilde{\beta} \alpha \, \Big| b_z + \omega_{\rmn{T}}^{-2} \bm b \bcdot \bnabla \theta \Big| }, \label{eq:thermal_diffusivity}
        \end{align}
        where $\theta$ is the Boussinesq buoyancy variable (proportional to temperature fluctuation, but with opposite sign; \citetalias{Perrone2022}), and $\alpha$ is a dimensionless parameter that absorbs all the information on the mean-free path and the typical cluster scales, which is defined as
    \begin{align}
        \alpha \equiv \frac{1}{\gamma} \left( \frac{\lambda_{\text{mfp,e}}}{H}\right)  \frac{H^2}{\chi_0 / \omega_{\rmn{T}}} , \label{eq:alpha_def}
    \end{align}
    where $\gamma$ is the adiabatic index and $H$ is the typical pressure scale-height of the cluster. By varying $\alpha$ within the range of realistic values expected in the ICM (see Section~\ref{appC}), we can simulate MTI turbulence  with stronger or weaker whistler suppression of thermal diffusion. Finally, the last quantity in the denominator of Eq.~\eqref{eq:thermal_diffusivity} represents the projection of the temperature gradient along the local direction of the magnetic field $\bm b$.
        Unless otherwise noted, in what follows we take $\omega_{\rmn{T}}^{-1}$ as the unit of time and  normalize all velocities to $\varv_{\mathrm{cond}}$. Similarly,  we non-dimensionalize $\theta$ as $\tilde{\theta} \equiv \theta / (N \varv_{\text{cond}})$.

        \subsection{Parameters in the periphery of galaxy clusters}\label{appC}

    To estimate the value of the $\alpha$ parameter in the periphery of galaxy clusters, we used Eq.~\eqref{eq:alpha_def} together with the definition of the Spitzer diffusivity and of the electron collisional mean-free path. This yields
    \begin{align}
        \alpha \simeq 0.01 \left( \frac{H}{300\,\si{kpc}} \right) \left( \frac{\omega_{\rmn{T}}}{(700 \, \si{Myr})^{-1}} \right) \left( \frac{T_\rmn{e}}{5 \, \si{keV}} \right)^{-1/2}, \label{eq:alpha_numbers}
    \end{align}
    where $H$, $\omega_{\rmn{T}}$, and $T_\rmn{e}$ are measured in units of $\si{kpc}$, $\si{Myr^{-1}}$, and $\si{keV}$, respectively. 
    It is important to note that these quantities can vary substantially both across different clusters and within the cluster periphery. Nevertheless, an order-of-magnitude estimate can be obtained assuming that clusters are in hydrostatic equilibrium, and taking power-law pressure and temperature profiles (justified in our local model). For the power-law exponents we take the best-fit profiles for the X-COP sample reported by \citet{Ghirardini2019}. With these assumptions, and taking a fiducial radius (where these quantities are computed) on the order of several hundred $\si{kpc}$, we find that the range of $\alpha$ allowed by observations is 0.01--0.05, where the lower estimate is obtained from Eq.~\eqref{eq:alpha_numbers} with the values given, while the higher estimate is calculated assuming $H = 600\,\si{kpc}$, $\omega_\rmn{T} = (600 \, \si{Myr})^{-1}$, and $T_\rmn{e} = 1 \, \si{keV}$.

        \section{Suppression of turbulence and demise of the MTI}
        
        We ran 2D and 3D simulations to study the impact of whistler-suppressed thermal conduction on the MTI (i.e., $\alpha\neq0)$ and  to compare the results against our reference run ($\alpha=0)$.  
        
    In Fig.~\ref{fig:whistler_suppression_1D_2D}b we show a snapshot of a representative 2D MTI run with whistler suppression at saturation. The simulation was initialized with a weak uniform magnetic field in the horizontal direction and no external forcing (see Appendix~\ref{appD} for further details). We note that despite the significant suppression of thermal diffusivity, the character of MTI turbulence remains qualitatively the same: hot plumes rise across the box, cold plumes sink. The shape of the plumes traces the morphology of the magnetic field.
        The volume average of $\chi / \chi_0$ is less than $20\%$, but the spatial distribution is highly inhomogeneous and correlates strongly with $\tilde{\beta}$. Conversely, the correlation with $L_{\rmn{T},\parallel}$ (not shown) is weaker.

 \begin{figure}
                \centering
                \includegraphics[width=1.0\columnwidth]{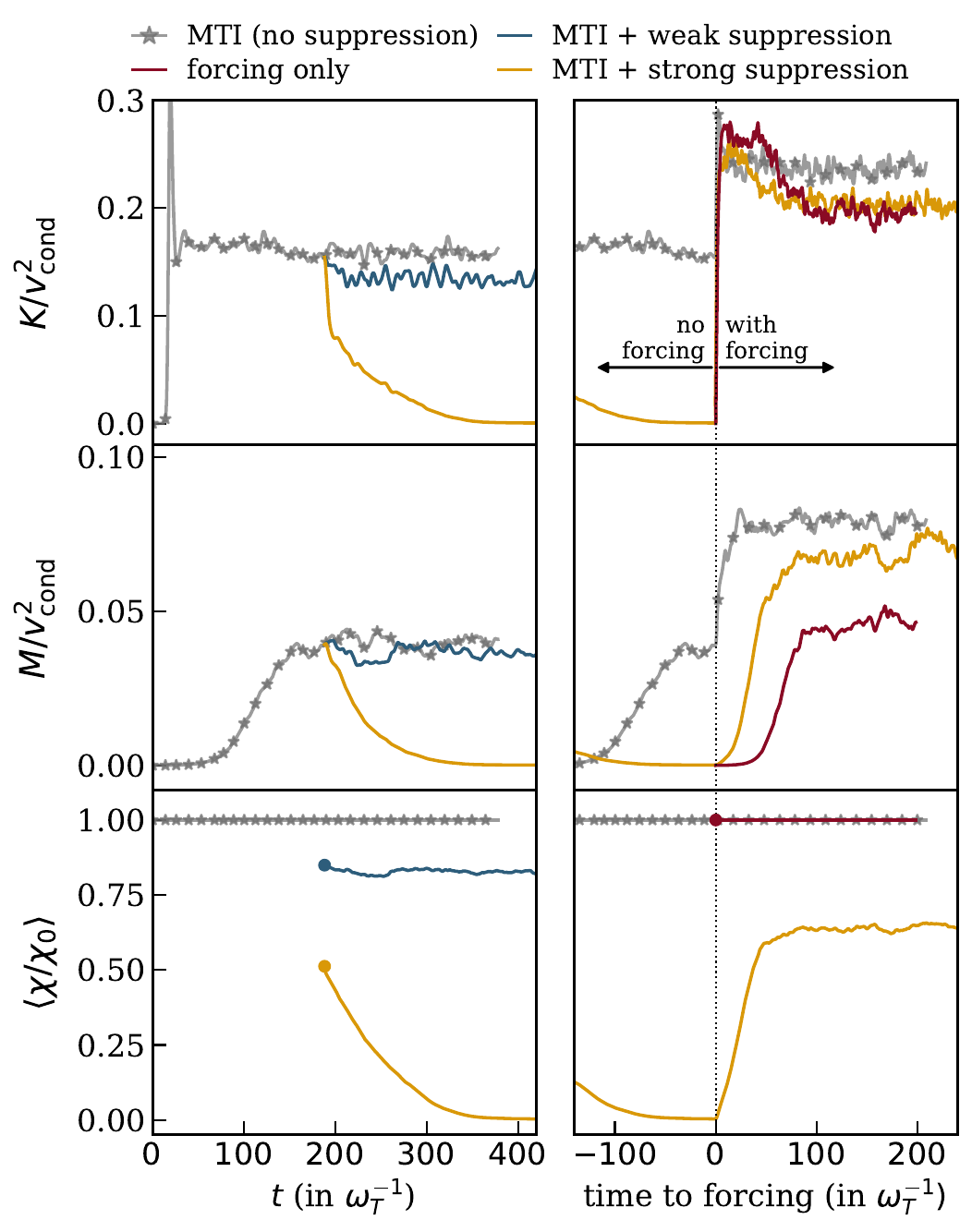}
                \caption{Time series of volume-averaged quantities in 3D MTI with whistler suppression. From top to bottom: Specific kinetic energy, specific magnetic energy, and thermal diffusivity. The left column shows our MTI runs without external forcing: without suppression (gray starred line), with weak suppression (solid blue), and with strong suppression (solid gold). In the right panel   external forcing is introduced. 
                For comparison,  also shown is a pure-forcing run (solid red line) with isotropic thermal diffusion (no MTI). MTI turbulence decays with strong whistler suppression, but the addition of forcing revives it.
                }
                \label{figure:volume-avg-forcing-suppression-onecol}%
        \end{figure}

    A limitation of 2D runs is that the magnetic field strength (and thus the overall suppression of thermal conduction) is somewhat arbitrary since they lack a small-scale dynamo that pins the magnetic energy at the equipartition value. As we  now show, in 3D the presence of the small-scale dynamo (or lack thereof) plays a crucial role in the whistler suppression of MTI turbulence. 
    In the left column of Fig.~\ref{figure:volume-avg-forcing-suppression-onecol} we compare the time evolution of the volume-averaged specific kinetic ($K \equiv \langle u^2 \rangle / 2$) and magnetic energy ($M \equiv \langle B^2 \rangle / (8 \pi \rho_0) $), as well as the suppression factor $\chi / \chi_0$ of a reference 3D MTI run ($\alpha=0$), and of two 3D runs with weak ($\alpha=0.012$) and strong ($\alpha=0.12$) suppression.
    We note that here ``weak'' and ``strong'' roughly correspond to the lower and higher ends of the range of whistler suppression that can be expected in the periphery of galaxy clusters (see Section~\ref{appC}). 
    As initial conditions for these latter runs, we used a turbulent snapshot of the reference run at saturation, and we then turned on whistler suppression (see Appendix~\ref{appD} for the numerical details). The initial spike in the kinetic energy for the reference MTI run results from the disruption of the fastest-growing MTI mode (seeded from the initial random perturbations) by parasitic Kelvin-Helmholtz instabilities \citepalias{Perrone2022a}. In Fig.~\ref{figure:volume-avg-forcing-suppression-onecol} the magnetic energy reaches approximate equipartition with the turbulent kinetic energy ($M \sim 0.25 K$). In terms of the modified plasma $\beta$ this approximately corresponds to $\tilde{\beta} \simeq 25$ (we note the equivalence $M / \varv_{\mathrm{cond}}^2 = \tilde{\beta}^{-1}$). Converting to the actual plasma $\beta$, this gives $\beta \sim 50$, depending on the values of $\varv_{\mathrm{cond}}$ and the sound speed, which is in the lower range of expected cluster values \citep[$\beta \sim 50 - 100$ ;][]{Carilli2002}. We note, however, that the exact value of the constant of proportionality (and thus of $\beta$) between kinetic and magnetic energies depends on the numerical parameters of the simulation.

    With weak whistler suppression, MTI turbulence carries on, although at lower levels proportional to the average level of suppression ($\simeq 82 \%$). Similarly, the magnetic energy saturates slightly below the reference run. With stronger suppression we witness a radically different scenario:  the kinetic and magnetic energies sharply decrease, with a near-total suppression of the thermal diffusivity. 
    The physical picture behind the demise of the MTI can be described as follows. Because of the relation of proportionality between thermal diffusivity and turbulent kinetic energy \citepalias{Perrone2022,Perrone2022a}, a strong initial suppression of $\chi$ (and therefore a reduction in $K$) means that magnetic fields cannot be maintained by the small-scale dynamo at the same equipartition level as before, and that they decay to adjust to the new turbulence levels. However, this decay leads to further suppression of the thermal diffusivity through its $\beta$-dependence, which in turns weakens the turbulence, and so on. The results of this self-reinforcing cycle is a runaway process that kills the MTI entirely. We note that this runaway process only occurs in the simulation with strong suppression ($\alpha=0.12$), while in our weak suppression run ($\alpha=0.012$) the system adjusts to the new equilibrium without any hint of runaway. 
    We attribute this behavior to the impairment of the small-scale dynamo, which for MTI turbulence requires a magnetic Reynolds number $\rmn{Rm}$ (defined as $\rmn{Rm} \equiv \rmn{Re} \times \rmn{Pm}$, where $\rmn{Re}$ is the Reynolds number at the integral scale and $\rmn{Pm}$ the ratio of the viscosity to magnetic resistivity) above a critical value of $\rmn{Rm}_\rmn{c} \simeq 35$ \citepalias{Perrone2022a}. 
    In our 3D reference run, $\rmn{Rm} \simeq 77$, which is well above the critical threshold, with $\rmn{Pm} = 4$ (see Appendix~\ref{appC} for further details on the numerics). In our weak suppression run the magnetic Reynolds number remains above the critical value throughout ($\rmn{Rm} \simeq 75$), while in our strong suppression run $\rmn{Rm}$ quickly drops below $\rmn{Rm_c}$ (around $t=230$ in Fig.~\ref{figure:volume-avg-forcing-suppression-onecol}) and the small-scale dynamo shuts down.
    The existence of a critical transition between the two outcomes is explored in detail in \citetalias{Perrone2023b}.

\section{Revival of the MTI by external turbulence}
        
        The results of the previous section may cast a cloud on the contribution of the MTI to turbulent driving in the periphery of galaxy clusters.
        However, other processes are expected to play a role, such as mergers and accretion events \citep[][]{Simionescu2019}. Even in the event of total suppression of the MTI, turbulence injected through these other processes can also amplify magnetic fields and, in so doing, restore thermal conductivity to a significant fraction of the Spitzer value. This could make it possible for the MTI to spring back into action. 
        
        To assess the feasibility of this scenario, we looked at the effect of external turbulence on the whistler-suppressed MTI by running a number of simulations with  solenoidal white-noise forcing, which is delta-correlated in time. We tuned energy injection by the external forcing so as to drive turbulence of comparable amplitude and at a scale similar to that of our reference MTI run. This is consistent with previous estimates of MTI turbulence in the periphery of galaxy clusters \citepalias{Perrone2022a}.
        
        We show in Fig.~\ref{figure:volume-avg-forcing-suppression-onecol} (right column) the kinetic energy, magnetic energy, and the mean thermal diffusivity of three representative 3D runs before and after enabling external forcing. These runs were chosen to illustrate the full range of possibilities between (i) reference MTI with the addition of external forcing and (ii) external forcing only with isotropic thermal conduction (and thus without the MTI). Between them is the MTI run with strong whistler suppression and forcing. The addition of forcing results in an increase in the turbulent kinetic energy of the reference MTI run and an even more significant increase in the magnetic energy. Similarly, compared to the forcing-only run, the presence of the MTI clearly shows up as stronger turbulent fluctuations.

    It is the run with strong whistler suppression that exhibits the most dramatic change. After turning on external forcing, the thermal diffusivity of this once-decaying run rapidly increases, and the turbulence reaches levels above those of the forcing-only run, a clear sign that the MTI has been revived. Eventually, the thermal diffusivity stabilizes at around $64 \% \chi_0$. 
    
\begin{figure}
                \centering
                \includegraphics[width=1.0\columnwidth]{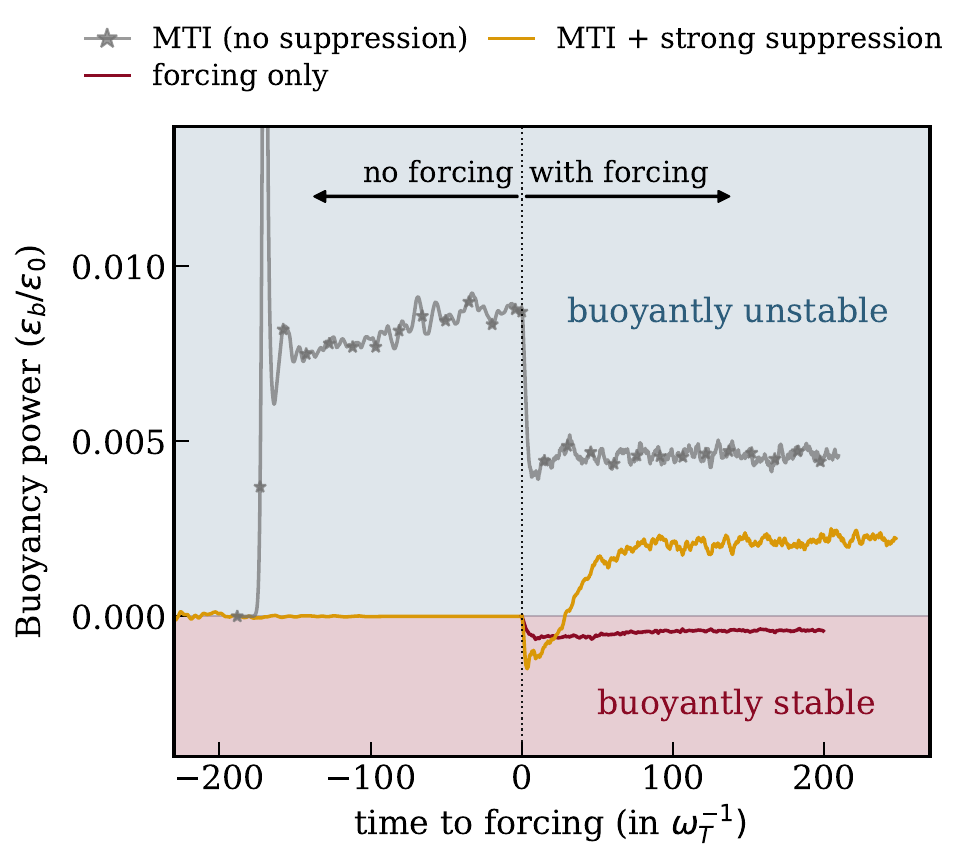}
                \caption{Evolution of the volume-averaged buoyancy power in 3D MTI with whistler suppression. To the left of the vertical dotted line the  runs have no additional external forcing, which is turned on at $t=0$. The buoyancy power is normalized to $\varepsilon_0 \equiv \chi_0 \omega_{\rmn{T}}^4 / N^2$, the MTI energy injection rate \citepalias{Perrone2022}. The 3D runs shown are the same as in Fig.~\ref{figure:volume-avg-forcing-suppression-onecol}. Despite strong whistler suppression, the system remains buoyantly unstable thanks to external forcing.}
                \label{figure:letter_forc_buoyancy_time}%
        \end{figure}

    The impact of external forcing on the MTI is particularly evident if we plot the time evolution of the volume-averaged buoyancy power $\varepsilon_\rmn{b} \equiv - \langle \theta u_z \rangle$ (Fig.~\ref{figure:letter_forc_buoyancy_time}),   the rate at which kinetic energy is added or removed by the buoyancy force, obtained by dotting the momentum equation (Eq.~\ref{eq:mom_eq}) with $\bm u$. As $\theta$ is proportional to the density fluctuation, the quantity is positive if there is an anticorrelation between $\theta$ and $u_z$, which is the case when lighter fluid travels upward and denser fluid sinks. A net-positive value of the buoyancy power thus indicates a buoyantly unstable flow, as in thermal convection and in the case of reference MTI (where $\varepsilon_\rmn{b}$ is proportional to the fraction of energy drawn from the background temperature gradient converted into velocity fluctuations by the MTI). Stably stratified flows instead have a net-negative buoyancy power, meaning that an initial perturbation will be damped.
    Whether the flow is buoyantly stable or unstable has important implications for the efficiency of plasma and metal mixing in the cool cores and in the bulk ICM \citep{Sharma2009b,Ruszkowski2010,Yang2016b,Kannan2017}.

    The addition of external turbulence clearly interferes with the MTI, which becomes less efficient at converting thermal energy into turbulence even in the case of no whistler suppression; in the reference MTI run $\varepsilon_\rmn{b}$ is reduced by about $50\%$ after turning on forcing. Remarkably, the run with strong whistler suppression goes from negligible energy injection from buoyancy before forcing (because of the decaying MTI) to significant energy injection  as the external forcing boosts thermal diffusion and revives the MTI (about one-half of the reference run with forcing, though the exact figure depends on the value of $\alpha$). This is to be contrasted with the pure-forcing run which has isotropic thermal diffusion, and therefore no MTI,  where the buoyancy power has an overall stabilizing effect.

        \begin{figure}
                \centering
                \includegraphics[width=1.0\columnwidth]{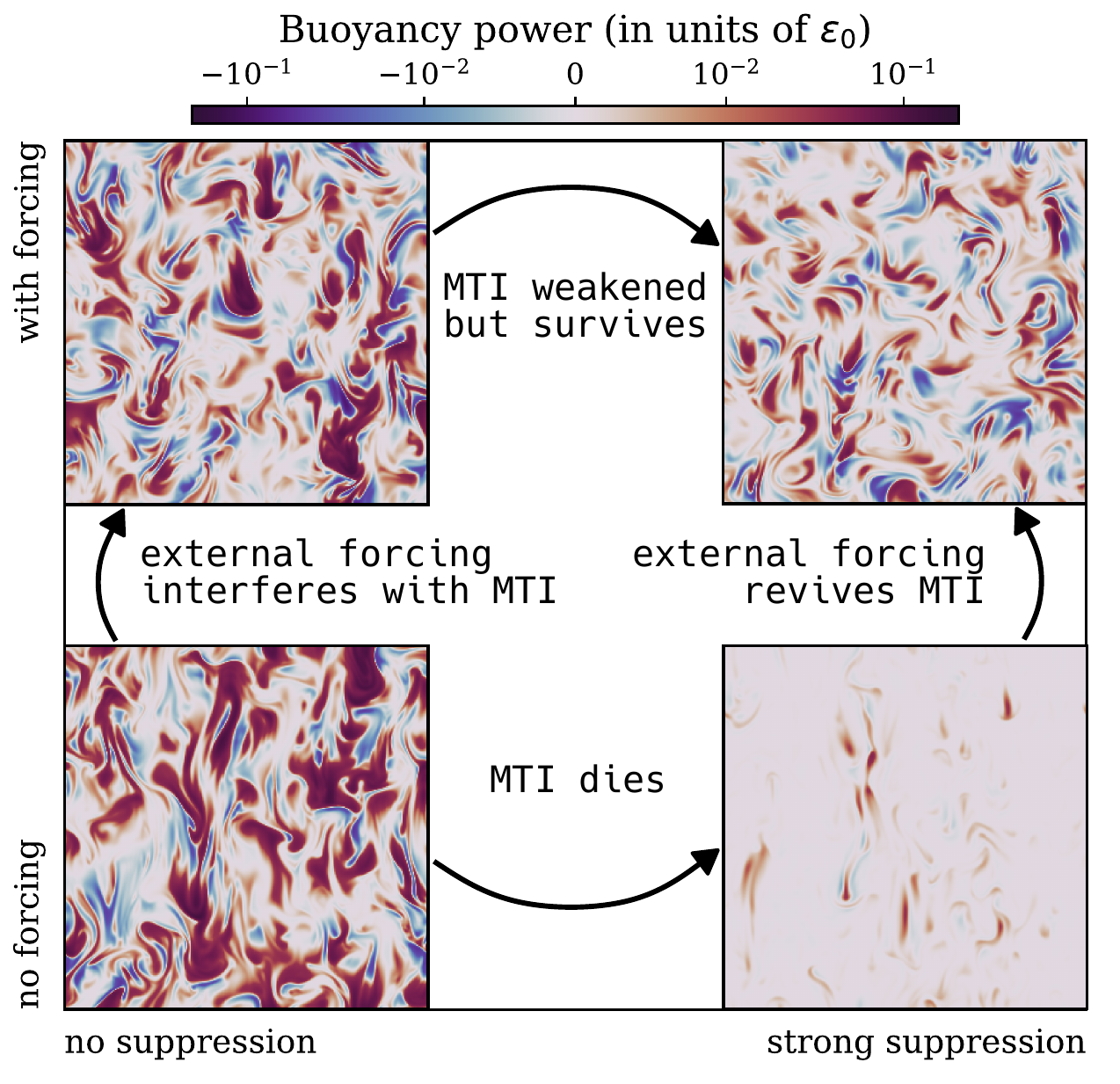}
                \caption{Buoyancy power for two 3D runs (reference MTI and strongly whistler-suppressed MTI) before and after turning on external forcing. The buoyancy power is in units of the MTI energy injection rate and its volume average is shown in Fig.~\ref{figure:letter_forc_buoyancy_time}. External forcing interferes with MTI turbulence, but the upwelling and downwelling of the plasma is still present (regions in red). }
                \label{figure:letter-cartoon}%
        \end{figure}
                
\section{Discussion and conclusions}

        In this work we explored the impact of whistler suppression of the heat flux on the MTI through small-scale simulations with a subgrid model for the thermal diffusivity inspired by kinetic theory and particle-in-cell simulations. Our main findings are that the impact can be more or less severe depending on the level of collisionality in the periphery of galaxy clusters. For weak whistler suppression, the main features of MTI-driven turbulence are qualitatively unchanged, with only an overall decrease in the turbulent fluctuations. For strong whistler suppression, the MTI loses the ability to drive a small-scale dynamo and decays. These two extremes span the range that can be realistically expected in the periphery of galaxy clusters, and it is thus not possible to rule out a priori whether the MTI would survive on its own.
        
        The MTI, however, is not the only source of turbulence in the bulk of the ICM. As we have shown,   and contrary to common intuition,  these other processes can have a beneficial effect on the MTI
        when thermal conduction is suppressed by the whistler instability.
        The basic mechanism is summarized in Fig.~\ref{figure:letter-cartoon} where we show snapshots of the buoyancy power for the runs without suppression ($\alpha = 0$) and  with strong suppression ($\alpha = 0.12$) before and after turning on external forcing. The bottom left corner shows the standard MTI without whistler suppression and external forcing. Turning on strong whistler suppression (bottom right) stalls the MTI-driven small-scale dynamo, and as a result  the turbulence dies out.
        On the other hand, external forcing without whistler suppression evidently interferes with the basic workings of the MTI (top left), but crucially the convective character of the instability is not erased. This leads to the main new finding of this work:     external turbulence is able to revive ``dead'' MTI simulations with strong whistler suppression, and that the flow remains overall buoyantly unstable (top right). In galaxy clusters  this change in buoyant response leads to an increased efficiency of turbulent mixing, which could have important astrophysical implications. For instance,  it could explain the increased metal mixing rates (and more efficient coupling of active galactic nuclei to the ambient ICM) that was found in cosmological simulations with anisotropic conduction \citep{Kannan2017,Barnes2019}.
    As we have shown,  even with whistler suppression  the convectively unstable nature of the ICM seems to be qualitatively unchanged.
    Our work illustrates the need for detailed modeling of the weakly collisional outskirts of galaxy clusters, and sets the stage for the development of effective models of turbulence in large-scale cosmological simulations.

        \begin{acknowledgements}
                The authors acknowledge support by the European Research Council under ERC-AdG grant PICOGAL-101019746. This project has received support from the European Union’s Horizon Europe research and innovation programme under the Marie Skłodowska-Curie grant agreement No 101106080.
        \end{acknowledgements}
        
        %
           \bibliographystyle{aa} 
        \bibliography{bibliography}
        %

        \begin{appendix} 
        \section{Methods}
    \label{app}

        \subsection{Boussinesq approximation to MHD}
    \label{appA}
     In the Boussinesq approximations we follow the evolution of small-scale, subsonic fluctuations within a stationary background in hydrostatic equilibrium (denoted by the subscript ``$0$'', e.g., $\rho_0$ for the mass density, $T_0$ for the temperature), which represents a local volume of a galaxy cluster at a reference radius $R_0$. The variables evolved by \textsc{snoopy} are the velocity fluctuation $\bm u$, the magnetic field $\bm B$ (rescaled by $\sqrt{4 \pi \rho_0}$ so that it has the dimensions of a velocity), and the buoyancy variable 
\begin{align}
    \theta \equiv g_0 \frac{\delta \rho}{\rho_0},
\end{align}     
     which is given in units of $\si{cm.s^{-2}}$, where $g_0$ is the gravitational acceleration. The appropriate equations are derived in \citetalias{Perrone2022}, and are shown below including a spatially varying anisotropic thermal diffusivity and an external acceleration $\bm f$ in the momentum equation:
\begin{align}
        \bnabla \bcdot \bm u = \bnabla \bcdot \bm B &= 0 ,\label{eq:div_eq} \\
        \left( \partial_t + \bm u \bcdot \bnabla \right) \bm u &= - \frac{\bnabla p_\text{tot}}{\rho_0}  - \theta \bm e_z + \left( \bm B \bcdot \bnabla \right) \bm B + \nu \bnabla^2 \bm u + \bm f, \label{eq:mom_eq} \\
        \left( \partial_t + \bm u \bcdot \bnabla \right) \bm B &= \left( \bm B \bcdot \bnabla \right) \bm u + \eta \bnabla^2 \bm B, \label{eq:bfield_eq} \\
        \left( \partial_t + \bm u \bcdot \bnabla \right) \theta &= N^2 u_z +  \bnabla \bcdot \left[ \chi \bm b \left( \bm b \bcdot \bnabla \right)\theta \right] +  \omega_{\rmn{T}}^2 \bnabla \bcdot \left( \chi \bm b b_z\right) \nonumber \\
        &~~~~- (-1)^n \nu_{\text{th}}^n \bnabla^{2n} \theta. \label{eq:buoyancy_eq}
\end{align}
    Here $p_\text{tot}$ is the sum of thermal and magnetic pressures; $\nu$ and $\eta$ are the fluid viscosity and magnetic resistivity, while in the buoyancy equation (Eq.~\ref{eq:buoyancy_eq}) the thermal diffusivity $\chi$ absorbs a factor of $(\gamma -1)/\gamma$. The (constant in $z$) squared MTI and Brunt-V\"ais\"al\"a frequencies that appear in  Eq.~\eqref{eq:buoyancy_eq} are defined through
        \begin{align}\label{eq:freq}
                N^2 &= \frac{g_0}{\gamma} \left. \frac{\partial (\ln p \rho^{-\gamma} ) }{\partial R} \right|_0 
        =\left.-\frac{1}{\gamma \rho}\frac{\partial \ln p}{\partial R}  \frac{\partial \ln (p \rho^{-\gamma})}{\partial R}\right|_0 ,\\
                \omega_{\rmn{T}}^2 &= \left.- g_0 \frac{\partial \ln T}{\partial R}\right|_0
        =\left.\frac{1}{\rho}\frac{\partial \ln p}{\partial R}  \frac{\partial \ln T}{\partial R}\right|_0,
        \end{align} 
    and are computed at the reference radius $R_0$. We note that in the buoyancy equation we have also included a higher-order diffusion operator (where $\nu_{\text{th}}^n$ denotes the corresponding hyperdiffusivity, not to be confused with $\nu$, the fluid viscosity) to regularize the smallest scales in our simulation in the event of strong spatial variations in the whistler-suppressed thermal diffusivity. We opted for a third-order ($n=3$) hyperdiffusion to allow for a wide scale separation between energy injection by the MTI and dissipation. The introduction of hyperdiffusivity results in a lower overall energy of the buoyancy fluctuations; for example, in our 3D reference run with hyperdiffusion the thermal energy is $\lesssim 7\%$ less than an equivalent run without hyperdiffusion, while the kinetic and magnetic energies are practically unchanged. However,  provided this suppression happens near the smallest scales of the box,  the resulting dynamics in spectral space are mostly unaffected.
    
    \subsection{Whistler-suppressed thermal conductivity in Boussinesq}\label{appB}
    
    The thermal diffusivity associated with Eq.~\eqref{eq:whistler_suppression_approx} is
    \begin{align}
    \chi = \frac{\chi_0}{1 + \frac{1}{3} \beta \lambda_{\text{mfp,e}}/L_{\rmn{T},\parallel}}. \label{eq:thermal_diffusivity_physical}
    \end{align}
    To rewrite it in terms of our Boussinesq variables, we first split the total temperature gradient parallel to the magnetic field $L_{\rmn{T},\parallel}$ into the contributions coming from the background profile plus the temperature perturbation
    \begin{align}
        L_{\rmn{T},\parallel}^{-1} &=  \left\lvert\frac{\bm b \bcdot \bnabla T}{T} \right\rvert \simeq  \left\lvert b_z \left( \frac{\mathrm{d} \ln T}{\mathrm{d} z} \right)_0  + (\bm b \bcdot \bnabla )\frac{\delta T}{T_0} \right\rvert         \nonumber \\
        &= \frac{\omega_{\rmn{T}}^2}{g_0} \left\lvert  b_z  + \omega_{\rmn{T}}^{-2} \bm b \bcdot \bnabla \theta \right\rvert. \label{eq:parallel_tempscale}
    \end{align}
    We also assume hydrostatic equilibrium to write the pressure scale-height as $H = c_{\rmn{s}}^2 / g_0$, where $c_{\rmn{s}}^2$ is the isothermal sound speed. Equation~\eqref{eq:thermal_diffusivity} is then obtained combining Eq.~\eqref{eq:parallel_tempscale} with the definitions for $\tilde{\beta}$ (Eq.~\ref{eq:beta_tilde_def}) and $\alpha$ (Eq.~\ref{eq:alpha_def}).

        \subsection{Initial conditions}\label{appD}
    We run numerical simulations in cubic (or square) boxes with an aspect ratio of unity. We show in this work a subset consisting of one 2D run with weak whistler suppression ($\alpha = 2 \times 10^{-3}$), and four 3D runs as follows: one reference MTI simulation ($\alpha = 0$), one MTI run with weak whistler suppression ($\alpha = 0.012$), one MTI run with strong whistler suppression ($\alpha = 0.12$), and one forcing-only run with isotropic thermal conduction and no whistler suppression ($\alpha = 0$). 
    The resolution is $1024^2$ for the 2D run, and $288^3$ for the 3D simulations. The parameters of the 2D run are $\chi_0 / (L^2 \omega_{\rmn{T}})  = 10^{-3}$, $\nu / \chi_0 = 0.01$, and the physical Prandtl number $\nu / \eta = 1$, while in 3D we choose $\chi_0 / (L^2 \omega_{\rmn{T}})  = 0.0025$, $\nu / \chi_0 = 0.042$, and $\nu / \eta = 4$; in the forcing-only run, we note that thermal conduction is isotropic with the same $\chi_0$. The hyperdiffusivity is $\nu_{\text{th}}^n / (\Delta h^4 \chi_0) = 5.2 \times 10^{-3}$ for the 2D run, and $\nu_{\text{th}}^n / (\Delta h^4 \chi_0) = 4.8 \times 10^{-3}$ for the 3D runs (with the exception of the forcing-only run, where it is zero), where $\Delta h$ is the grid size (uniform in all directions). All our runs have $N^2/ \omega_{\rmn{T}}^2 = 0.1$. 
    
    The 2D run, the reference 3D MTI run, and the 3D forcing-only run are initialized with random velocity fluctuations of amplitude $|\delta u_i| / (L \omega_{\rmn{T}}) = 10^{-4} $, for $i=1,2,3$. While the 2D run has a uniform horizontal magnetic field (with initial $B_0/ (L \omega_{\rmn{T}}) = 10^{-4}$), the reference 3D MTI run and forcing-only have no-net magnetic flux, and the initial magnetic field is given by $\bm B = B_0 [  \sin(2 \pi z / L)\bm e_x + \cos(2 \pi z / L)\bm e_y ]$, with the same $B_0$ as in 2D.
    These values correspond to $\tilde{\beta}_0 = 2 \times 10^5$ in 2D and $\tilde{\beta}_0 = 5 \times 10^5$ in 3D (volume-averaged).
    
    The 3D MTI runs with weak and strong whistler suppression instead use as initial conditions a snapshot of the 3D MTI reference run. Similarly, when we turn on forcing we use as initial condition a snapshot of the corresponding run without forcing.

        \end{appendix}

\end{document}